\documentclass{article}
\usepackage{bm,latexsym,amsmath,amssymb,amsfonts,fancyhdr,color,graphicx,multirow,slashed,cite}
\usepackage[a4paper,bottom=3cm,top=2.5cm,head=0mm,width=17cm,dvipdfm]{geometry}
\usepackage[usenames,dvipsnames,svgnames,table]{xcolor}
\usepackage[colorlinks=true,
            linkcolor=blue,
            urlcolor=blue,
            citecolor=green,          
						bookmarks=true,
						bookmarksnumbered=true,
						breaklinks=true,
						pdfpagemode=Fullscreen,
						pdfstartview=FitBH]{hyperref}
\usepackage[dotinlabels]{titletoc}
\usepackage{titlesec}
\usepackage{authblk,ulem}

\usepackage{multirow}

\numberwithin{equation}{section}
\allowdisplaybreaks[4]

\titlelabel{\thetitle.\quad \hspace{-0.8em}}
\titlecontents{section}
              [1.5em]
              {\vspace{4mm} \large \bf}
              {\contentslabel{1em}}
              {\hspace*{-1em}}
              {\titlerule*[.5pc]{.}\contentspage}
\titlecontents{subsection}
              [3.5em]
              {\vspace{2mm}}
              {\contentslabel{1.8em}}
              {\hspace*{.3em}}
              {\titlerule*[.5pc]{.}\contentspage}
\titlecontents{subsubsection}
              [5.5em]
              {\vspace{2mm}}
              {\contentslabel{2.5em}}
              {\hspace*{.3em}}
              {\titlerule*[.5pc]{.}\contentspage}

\newcommand{\titledef}{MeV Electrophilic Axion-like Particles from Sun} 

\hypersetup{ pdfauthor = {Shao-Feng Ge},
	     pdftitle = {\titledef}, 
	     pdfsubject = {}, 
             pdfkeywords = {}, 
	     pdfcreator = {LaTeX with hyperref package},
	     pdfproducer = {dvips + ps2pdf} }

\definecolor{gesfblack}{rgb}{0,0,0}

\definecolor{gesfblue}{rgb}{0.08,0.42,0.76}

\definecolor{gesfgreen}{rgb}{0,1,0}

\definecolor{gesfgrey}{rgb}{0.5,0.5,0.5}

\definecolor{gesflanse}{rgb}{0.00,0.50,0.50}

\definecolor{gesfpurple}{rgb}{0.47,0.19,0.42}

\definecolor{gesfred}{rgb}{1,0,0}

\definecolor{gesfwhite}{rgb}{1,1,1}

\definecolor{gesfyellow}{rgb}{0.7,0.4,0.3}

\newcommand{\gsec}[1]{{\hypersetup{linkcolor=red}Sec.\,\ref{#1}\hypersetup{linkcolor=blue}}}

\newcommand{\geqn}[1]{\hypersetup{linkcolor=blue}Eq.\,(\ref{#1})\hypersetup{linkcolor=blue}}
\newcommand{\gfig}[1]{{\hypersetup{linkcolor=violet}Fig.\,\ref{#1}\hypersetup{linkcolor=blue}}}

\definecolor{Orange}{cmyk}{0,0.61,0.87,0}
\definecolor{JungleGreen}{cmyk}{0.99,0,0.52,0}
\definecolor{OliveGreen}{cmyk}{0.64,0,0.95,0.40}
\definecolor{Brown}{cmyk}{0,0.81,1,0.60}
\definecolor{RoyalBlue}{cmyk}{0.71,0.53,0,0.12}
\definecolor{Gray}{cmyk}{0,0,0,0.40}
\definecolor{LightPink}{cmyk}{0.0,0.25,0,0}
\definecolor{LLightPink}{cmyk}{0.0,0.10,0,0}
\definecolor{LightBlue}{cmyk}{0.25,0,0,0}
\definecolor{LightGray}{cmyk}{0,0,0,0.2}

\graphicspath{{figs/}}

\begin{document}
\fontsize{12pt}{14pt}\selectfont

\title{
       \textbf{\fontsize{19pt}{21pt}\selectfont \titledef}} 
\author[1,2]{{\large Shao-Feng Ge} \footnote{\href{mailto:gesf@sjtu.edu.cn}{gesf@sjtu.edu.cn}}}
\author[1,2]{{\large Sk Jeesun} \footnote{\href{mailto:jeesun@sjtu.edu.cn}{jeesun@sjtu.edu.cn}}}
\author[3,4]{{\large Tao Li} \footnote{\href{mailto:taoli@sjtu.edu.cn}{taoli@sjtu.edu.cn}}}
\affil[1]{State Key Laboratory of Dark Matter Physics, Tsung-Dao Lee Institute \& School of Physics and Astronomy, Shanghai Jiao Tong University, Shanghai 200240, China}
\affil[2]{Key Laboratory for Particle Astrophysics and Cosmology (MOE) \& Shanghai Key Laboratory for Particle Physics and Cosmology, Shanghai Jiao Tong University, Shanghai 200240, China}
\affil[3]{SJTU Paris Elite Institute of Technology, Shanghai Jiao Tong University, Shanghai 200240, China}
\affil[4]{Shanghai Jiao Tong University Sichuan Research Institute, Chengdu 610213, China}
\date{\today}

\maketitle

\begin{abstract}
This work explores the production of an MeV-scale
electrophilic axion-like particles (ALPs) by utilizing the monochromatic
5.5\,MeV photon resulting from the nuclear
fusion processes in the Sun. 
These 5.5 MeV photons can undergo the Compton-like scattering
with the ambient electrons in the solar matter to produce
a substantial flux of MeV ALPs. 
Upon reaching the Earth, such ALPs can be detected via the same electron coupling,
offering a new opportunity for the dark matter (DM)
direct detection experiments to probe the previously
unexplored parameter regions.
We show that the existing data of LZ, PandaX-4T, and Borexino
can attain the sensitivities $g_{ae} \lesssim 3.7 \times 10^{-6}$, $g_{ae} \lesssim 3.7 \times 10^{-6}$ and
$g_{ae} \lesssim 1.7 \times 10^{-6}$, respectively,
for $m_a\lesssim1$\,MeV.
An optimistic 200 tonne$\times$year exposure by PandaX-xT
can reach $g_{ae}\lesssim 1.6 \times 10^{-6}$ for most
of the mass window $m_a < 1$\,MeV and even
$g_{ae} \lesssim 1.5 \times 10^{-7}$ with $m_a$
approaching 1\,MeV.
Despite the stringent constraints from different laboratory
experiments and astrophysical observations, our obtained 
limits from LZ, PandaX-4T, and Borexino can probe new parameter regions, specifically in the mass window $0.4\,{\rm MeV}\lesssim m_a\lesssim 1$\,MeV.
\end{abstract}

\section{Introduction}
%

The Standard Model (SM) of particle physics has
been remarkably successful in describing the
fundamental interactions of nature. Nevertheless,
it falls short in addressing several profound
puzzles of the Universe including the strong CP problem
\cite{ParticleDataGroup:2024cfk}.
To resolve these issues, SM is often augmented
with one or more beyond the Standard Model (BSM) particles.
Axion is one of such well motivated and widely explored BSM
extensions which was originally coined to solve
the strong CP problem in the quantum chromodynamics
(QCD) sector \cite{Peccei:1977hh,Wilczek:1977pj}.
Axion emerges as a pseudo-Nambu-Goldstone boson (pNGB) from a spontaneously broken global $U(1)$ symmetry often referred as Peccei-Quinn (PQ) symmetry.
In such a framework, the mass of Axion is inversely proportional to the symmetry breaking scale ($f_a$) which is typically much higher than the electro-weak (EW) scale.   
Several well established ultraviolet (UV) complete
models with such an axion such as Peccei–Quinn–Weinberg–Wilczek
(PQWW) model \cite{Peccei:1977hh,Weinberg:1977ma,Wilczek:1977pj},
the Kim–Shifman–Vainshtein–Zakharov (KSVZ) model
\cite{Kim:1979if,Shifman:1979if}, and the
Dine–Fischler–Srednicki–Zhitnitsky (DFSZ) model
\cite{Zhitnitsky:1980tq,Dine:1981rt} have been proposed.
Since the QCD axion mass and couplings are correlated,
some of the QCD models are strongly constrained
\cite{Peccei:2006as,DiLuzio:2020wdo}.
With such mass and coupling correlation relaxed,
the axion-like particle (ALP) is broadly used
to denote those pseudo-Nambu–Goldstone bosons that
are not necessarily tied to
the strong CP problem and can emerge in a wide
class of BSM scenarios featuring spontaneously
broken global $U(1)$ symmetries at some higher
energy ($\gg$ EW) scales
\cite{Biekotter:2025fll,DiLuzio:2020wdo,Bharucha:2022lty}.
In other words, ALPs exhibit a more flexible parameter space,
with masses and couplings being largely independent.

Axion and ALPs can play diverse roles in particle physics and cosmology, including viable dark matter candidates \cite{Dine:1982ah,Preskill:1982cy, 
Abbott:1982af, DiLuzio:2020wdo, Co:2019jts,Arias:2012az,Jaeckel:2014qea}, mediators between visible and hidden sectors \cite{Bharucha:2022lty,Ghosh:2023tyz,Fiorentino:2026dea}, and probes of high-scale BSM physics \cite{Jaeckel:2010ni,deGiorgi:2025ldc}. 
The rich phenomenological implications of interactions
between ALP and SM particles have motivated significant
theoretical and experimental investigations across multiple
frontiers, including laboratory searches \cite{Bauer:2017ris},
astrophysical observations \cite{Caputo:2024oqc}, and
cosmological probes \cite{Marsh:2015xka,Caloni:2022uya,Cadamuro:2010cz}.
Colliders with TeV center-of-mass energy range provide
the strongest constraints for heavier ALPs with masses
in the GeV$\sim$TeV range
\cite{Jaeckel:2015jla,Bauer:2017ris,Adhikary:2024mzi,BESIII:2022rzz}.
In contrast, experiments such as BaBar \cite{BaBar:2014zli}
and Belle  \cite{Inguglia:2016acz} impose the strongest
constraint for ALP with masses in the range
$\mathcal{O}(10^{-1})\,{\rm GeV} \sim \mathcal{O}(10)$\,GeV mass range.
Meanwhile, the beam dump experiments \cite{CHARM:1985anb, Bjorken:1988as,NA64:2020qwq,Afik:2023mhj,Ema:2023tjg,Pustyntsev:2026znq} have placed stringent constraints on ALPs in the $1 \sim 100$\,MeV mass window. 
On the astrophysical side, ALP interactions with quarks, electrons, and photons can lead to nontrivial consequences in stellar environments. 
The light and weakly coupled ALPs can be efficiently produced in stellar bodies 
to enhance energy loss and accelerate stellar cooling \cite{Vysotsky:1978dc,Giannotti:2015kwo}.
Among the cooling constraints, supernova with its huge density of SM particles places the strongest limit on ALP coupling up to ALP mass $\mathcal{O}(100)$\,MeV \cite{Carenza:2021pcm,Ferreira:2022xlw,Carenza:2023lci,Lella:2023bfb}.
In addition, the supernova ALPs can
also decay to photons and charged leptons which
would deposit energies in the environment to form
fireballs \cite{Diamond:2023scc,Fiorillo:2025yzf,Candon:2025ypl}.
Similar approach can also apply to neutron star mergers \cite{Diamond:2023cto}.
The supernova explosions can also emit ALPs that propagate to Earth, where they may be detected in multi-ton neutrino detectors \cite{Lella:2022uwi,Lucente:2022vuo,Alonso-Gonzalez:2024ems,Alonso-Gonzalez:2024spi,Alonso-Gonzalez:2026syw}.
On the cosmological side, ALPs with mass $\lesssim \mathcal{O}(1)$\,MeV
can potentially affect the effective relativistic degrees
of freedom \cite{Barbieri:2026ewj,Ghosh:2020vti} or primordial
abundances during the big bang nucleosynthesis (BBN)
\cite{Depta:2020wmr,Escudero:2025avx} and thereby face
strong constraints from cosmological observations.

Sun can also be extremely helpful to probe such feebly interacting BSM particles.
Using the large SM density and its thermal energy, the
Sun can produce a significant amount of keV ALPs. 
The production of the photophilic or electrophilic ALP through
the  Primakoff
\cite{Wu:2024fsf}
or Compton-like scattering, and bremsstrahlung
\cite{Derbin:2011gg,Carenza:2024ehj} processes
have already been explored in the existing literature \cite{Redondo:2013wwa,Hoof:2021mld}.
Similar processes can also apply to dark photons
\cite{Redondo:2013lna,Fabbrichesi:2020wbt,An:2020bxd,Antel:2023hkf} and
milli-charged particles \cite{Davidson:2000hf}.
Most of these emitted BSM particles have energies
$\sim$\,keV that corresponds to the average
temperature inside the Sun.
Thanks to the feeble coupling, such ALPs can escape the Sun
and reach the ground-based direct detection experiments, offering a unique opportunity to directly probe them.  

It is also possible to go beyond the thermal energy limitation
inside the Sun. 
The solar energies are originally produced from the
nuclear fusion processes, including both the $pp$
chain and the CNO circle. 
As one may expect, a nuclear process
can typically have much higher energy than $\mathcal O(1)$\,keV.
In particular, the $^2$H$ + p \rightarrow ^3$He$ + \gamma$
process can release a gamma photon with energy 5.5\,MeV
\cite{Vinyoles:2016djt}.
By coupling with nucleons, axion or ALP can
replace the gamma photon and be
directly produced from the nuclear process
$^2$H$ + p \rightarrow ^3$He$ + a$ 
\cite{CAST:2009klq,Derbin:2010dv,Borexino:2012guz,Bhusal:2020bvx,Vergados:2021ejk,Lucente:2022esm,Arias-Aragon:2024gdz,Haxton:2026ura,Haxton:2026pty}.
Besides, Sun can produce a mono-energetic axion from
the M1 transition of $^{57}$Fe (14.4\,keV)
\cite{Moriyama:1995bz,Krcmar:1998xn,Namba:2007rm,Derbin:2007hc,Derbin:2009yb,CAST:2009jdc,Derbin:2011zz,CUORE:2012ymr},
$^7$Li (478\,keV) \cite{Krcmar:2001si,Derbin:2005xc,Borexino:2008wiu,Belli:2008zzb,CAST:2009klq,Belli:2012zz},
or $^{83}$Kr (9.4\,keV) \cite{Jakovcic:2004sh,Gavrilyuk:2014mch}.
In addition, the mono-energetic gamma can be replaced
by a pair of fermions such that
nucleon-philic dark fermions can also be produced
through the nuclear fusions inside the solar core
\cite{Ge:2024cto}. 
Note that these production mechanisms
require direct couplings of ALP with nucleons. 

In this work, we explore an alternative scenario where
the ALP exclusively couples with electron such that it
can be produced with the $\mathcal O($MeV) nuclear energy.
Instead of replacing the 5.5\,MeV photon in the $pp$ chain,
an electrophilic ALP can be produced from the Compton-like
scattering when the energetic photon goes through the solar
medium. We first discuss the production and expected flux
of such MeV solar ALP in \gsec{sec:production} and then
studies its detection prospect at DM detectors in
\gsec{sec:detection}. Our conclusions and discussions
can be found in \gsec{sec:concl}.

\section{Solar Production with Electron Target}
\label{sec:production}
As mentioned in the Introduction, the solar energy
can be used to produce energetic axion-like particles
to overcome the detection thresholds. The major flux of electrophilic
solar axions is produced from the atomic,
bremstrahlung, and Compton-like processes
\cite{Redondo:2013wwa,Carenza:2024ehj}.
All these three production mechanisms essentially use
the thermal energy which can reach 10\,keV
around the solar core. However, this is the highest
energy that the Sun can provide.
While the solar thermal energy is typically in the
keV range, the nuclear energy
can reach $\mathcal O$(MeV). Especially,
the $^2$H$ + p \rightarrow ^3$He$ + \gamma$ process
can release a 5.5\,MeV $\gamma$ photon
\cite{Vinyoles:2016djt,Bhusal:2020bvx,Vergados:2021ejk,Lucente:2022esm,Arias-Aragon:2024gdz}. It is then a perfect
place to produce MeV mass dark sector particles.
Besides, there are several other channels
of producing axions with nuclear coupling e.g. from
the M1 transition of $^{57}$Fe (14.4\,keV)
\cite{Moriyama:1995bz,Krcmar:1998xn,Namba:2007rm,Derbin:2007hc,Derbin:2009yb,CAST:2009jdc,Derbin:2011zz,CUORE:2012ymr},
$^7$Li (478\,keV) \cite{Krcmar:2001si,Derbin:2005xc,Borexino:2008wiu,Belli:2008zzb,CAST:2009klq,Belli:2012zz},
or $^{83}$Kr (9.4\,keV) \cite{Jakovcic:2004sh,Gavrilyuk:2014mch}.
However, this requires the dark sector particles
to couple with nucleons in the first place.

We would like to point out that an electrophilic
dark sector particle can also be produced out
of the MeV nuclear energy. Once the 5.5\,MeV $\gamma$
photon is released from the proton and deutron
fusion, it would scatter with the electrons
around. This allows a Compton-like
process, $\gamma + e \rightarrow e + X$ where
$X$ represents any other final-state particles
besides the electron $e$.

In this paper, we work with an ALP $a$ that
couples to only the electron,
\begin{eqnarray}
    \mathcal{L}
= \frac{1}{2} \partial^\mu a \partial_\mu a
- \frac{1}{2}m_a^2 a^2 -i g_{ae}a\Bar{e}\gamma^5 e,
    \label{eq:lag}
\end{eqnarray}
where $m_a$ and $g_{ae}$ are the ALP mass and coupling with electron ($e$).
As mentioned in the Introduction, ALPs can be realized as a generic pNGBs originated from the spontaneous breaking of any global $U(1)$ symmetry existing at a higher energy scale.
Typically, this scale is assumed to be much larger than the characteristic energy scale ($\mathcal{O}(1)$\,MeV) of the processes considered in this work.
In this regime, the interactions of ALPs with Standard Model (SM) particles can be described within an effective field theory (EFT) framework \cite{Bauer:2017ris,DiLuzio:2020wdo}.
There also exist multiple ultraviolet (UV) complete models \cite{Dine:1981rt,Bharucha:2022lty}, which can lead to effective dimension-five operators exhibiting derivative couplings between the ALP and the axial current of SM fermions. Using the Dirac equation together with a suitable field redefinition, this derivative interaction can be rewritten in the pseudo-Yukawa form shown in \geqn{eq:lag}.

\begin{figure}[t]
\centering
\includegraphics[width=0.36\linewidth]{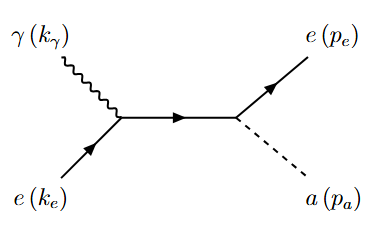}
\qquad
\includegraphics[width=0.36\linewidth]{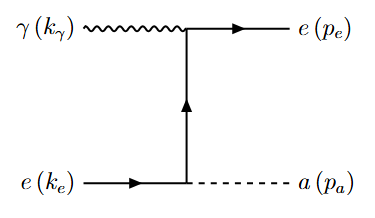}
\caption{Feynman diagrams for the solar production of
ALP $a$ through the Compton-like scattering
$\gamma + e \rightarrow a + e$
of a 5.5\,MeV photon and an electron.}
\label{fig:fd}
\end{figure}

With such interactions, ALPs can be produced copiously
in the Sun due to its huge volume.
The previously mentioned 5.5\,MeV gamma photon from
$p + d \rightarrow {}^3{\rm He} + \gamma$ \cite{BOREXINO:2018ohr}
can experience the following Compton-like scattering,
\begin{eqnarray}
  \gamma (k_\gamma) + e^-(k_e)
\to
  e^-(p_e) + a (p_a) ,
  \label{eq:prod}
\end{eqnarray}
to produce an energetic ALP $a$.
Those labels in the parenthesis signify the
respective momenta with $k$ for the initial-state
particles and $p$ for the final ones.
The Feynman diagrams relevant for ALP production
are shown in \gfig{fig:fd}.

Similar as the usual Compton scattering process,
there are two diagrams with the intermediate
electron in the $s$ or $t$ channel.
We denote the amplitudes for $s$
(left panel of \gfig{fig:fd}) and $t$ (right panel)
channel diagrams as $\mathcal M_s$
and $\mathcal M_t$ respectively.
The total amplitude
$\mathcal M \equiv \mathcal M_s + \mathcal M_t$
gives the spin averaged,
\begin{align}
  \overline{|\mathcal{M}|^2}
= \, &
  \frac{e^2 g_{ae}^2}{2 E_{\gamma}^2 m_e \left( m_a^2 - 2 E_a m_e \right)^2}
\left\{
-E_{\gamma} m_a^6 + m_a^4 \left( 2 E_a E_{\gamma} + m_a^2 \right) m_e
\right.
\nonumber
\\
+ & \, 4 \left( -E_a + E_{\gamma} \right) m_a^2
\left[ \left( E_a - E_{\gamma} \right) E_{\gamma}+ m_a^2 \right] m_e^2
\nonumber
\\
+ &
\left.
  4 \left( E_a - E_{\gamma} \right)^2 \left( 2 E_a E_{\gamma} + m_a^2 \right) m_e^3
\right\},
\label{eq:ampP}
\end{align}
where $E_\gamma$ and $E_a$ are the energy of the incoming photon and outgoing ALP, respectively, while $m_e$
is the electron mass.
Since the incoming photon has $\mathcal{O}(1)$\,MeV
energy which is much larger than the typical solar
electron kinetic energy, we treat the initial-state electron
as at rest.

In the rest frame of the target electron, the differential cross section with respect to the ALP energy takes the form,
\begin{eqnarray}
  \frac{d\sigma(E_\gamma)}{dE_a}
= \frac{1}{4 E_\gamma m_e v_{\rm rel}} \frac{1}{8\pi |\bm p_\gamma|}~\overline{|\mathcal{M}|^2},
\end{eqnarray}
with $|\bm p_\gamma|$ being the magnitude of the
3-momentum of the incoming photon and
$v_{\rm rel} = 1$ being the relative velocity which
is the speed of light for an incoming photon.
Note that the kinematically allowed energy range of
the outgoing ALP is also determined by the energy–momentum
conservation for a $2\to2$ scattering process in the rest
frame,
\begin{align}
  E_a^{\rm min,~max}
& =
  \frac 1 {2(m_e^2+2 m_e E_\gamma)}
\bigg\{ 
    (2 m_e E_\gamma+m_a^2)(E_\gamma+m_e)
\nonumber
\\
& 
  \mp E_\gamma
  \sqrt{\left[ m_e^2+2 m_e E_\gamma-(m_a+m_e)^2 \right]
        \left[ m_e^2+2 m_e E_\gamma-(m_a-m_e)^2 \right]}
\bigg\},
\label{eq:emin}
\end{align}
where the $-(+)$ sign is applicable for $E_a^{\rm min}$ ($E_a^{\rm max}$), respectively.

The expected solar ALP flux in our scenario can be
obtained from the previously quoted differential
cross section $d \sigma / d E_a$ by imposing the
energy limits,
\begin{eqnarray}
  \frac{d\Phi_a(E_a)}{dE_a}
= \frac{1}{4\pi d_{\odot}^2}
  \int d E_\gamma
  \frac{dN_\gamma}{dE_\gamma}
  \frac{1}{\sigma_{\rm tot}}
  \frac{d\sigma}{dE_a}~\Theta(E_a-E_a^{\rm min})\Theta(E_a^{\rm max}-E_a),
  \label{eq:fl}
\end{eqnarray}
where $d N_\gamma / dE_\gamma$ is the photon energy spectrum
per second.
For our scenario, the incoming energetic photon spectra is a monochromatic 
flux at energy 5.5\,MeV, 
$d N_\gamma / dE_\gamma = N_\gamma~\delta(E_\gamma-E_0)$ where $E_0=5.5\,{\rm MeV}$.
These gamma photons are produced through the nuclear
interaction $p+D\to ^3He + \gamma$ where $99.6\%$ of
the $D$ are produced from the $p-p$ chain
(e.g. $p+p\to D+ e^+ +\nu$). In other words, the
number $N_\gamma$ of these gamma photons can be estimated
from the same as the number of $pp$ neutrinos
\cite{Pospelov:2017kep}. This $\nu$ flux,
$\Phi_{pp\nu}$ ($\equiv N_{pp\nu} / 4 \pi d_\odot^2$
with $N_{pp\nu}$ being the number of $pp$ neutrinos
produced per second and $d_{\odot}=1.48\times 10^8$\,km
the distance from Sun to Earth) has been measured by
the Borexino experiment \cite{BOREXINO:2018ohr}
and can be estimated to be
$\Phi_{pp\nu}=10^{11}{\rm~cm}^{-2}{\rm s}^{-1}$.
For simple estimation, the number of 5.5\,MeV photons
$N_\gamma$ is roughly $4 \pi d_\odot^2 \Phi_{pp\nu}$.

Once the monochromatic gamma photon is injected into the
solar medium, it can experience either the QED Compton
scattering $\gamma+e\to\gamma+e$
or the Compton-like process for producing
an ALP. Since the Compton-like process is expected to
contribute only a small fraction of all the processes
that a gamma photon can experience, the produced ALP
flux scales with the differential fraction,
$d \sigma / \sigma_{\rm tot} d E_a$ where
$\sigma_{\rm tot}$ is the total cross section.
We take the QED Compton scattering cross section
$\sigma_E$ as $8\pi \alpha^2/m_e^2$ where $\alpha$ is
the fine-structure constant.

\begin{figure}[t]
\centering
\includegraphics[width=0.68\textwidth]{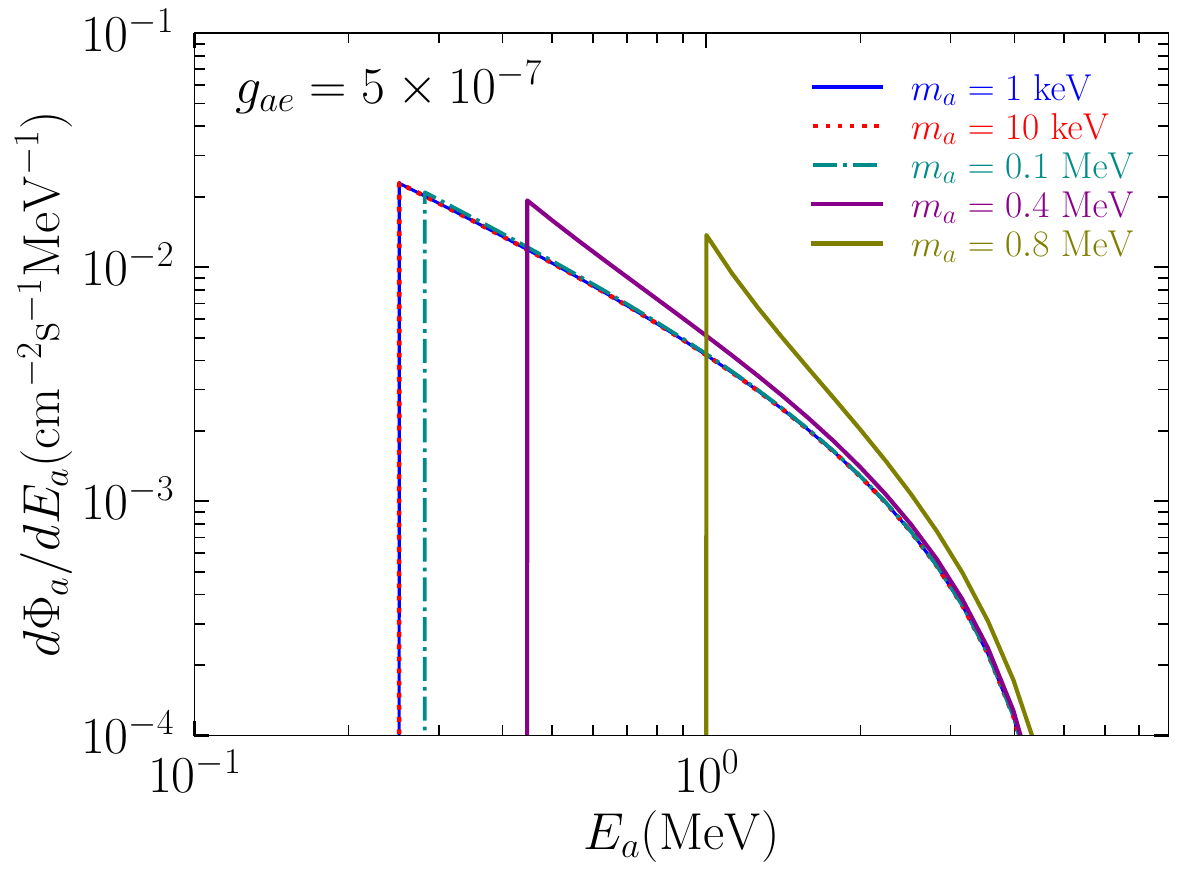}
\caption{The MeV solar axion differential flux arriving
at Earth with $g_{ae}=5 \times 10^{-7}$  for different
axion masses $m_a=1\,{\rm keV}$ (blue solid),
10\,keV (red dotted), 0.1\,MeV (dark cyan dash-dotted),
0.4\,MeV (magenta solid), and 0.8\,MeV (olive solid curves),
respectively.}
\label{fig:flux}
\end{figure}
The differential flux of solar ALP $a$ at Earth as
obtained from \geqn{eq:fl} is shown in \gfig{fig:flux}
for a fixed coupling constant $g_{ae}=5 \times 10^{-7}$.
The fluxes obtained with different masses $m_a=1\,{\rm keV}$,
10\,keV, 0.1\,MeV, 0.4\,MeV, 0.8\,MeV have been denoted by
blue solid, red dotted, dark cyan dash-dotted, magenta solid,
and olive solid curves, respectively.
The lower cut-off in the flux indicates the minimum
energy carried away by ALP according to \geqn{eq:emin}.  
For a vanishingly small axion mass, $m_a \ll m_e \ll E_\gamma$,
the minimum energy reads as $E_a^{\rm min} \approx m_e/2$. 
For higher ALP masses, the energy lower limit $E_a^{\rm min}$
also increases.
The decrease in the flux with an increasing $E_a$ can
also be understood by analyzing the amplitude given in
\geqn{eq:ampP}.
In the regime $m_a\ll m_e, E_\gamma$ limit, the scattering
amplitude $\overline{|\mathcal M|^2}$ in \geqn{eq:ampP}
simplifies to $2(E_a-E_\gamma)^2/E_a E_\gamma$, which
accounts for the observed decreasing behavior of the flux with an increasing $E_a$.
For all cases, the spectrum shrinks to a small enough value
around 4\,MeV since the Compton-like process typically
transfers only part of the incoming photon energy.
However, such energy is large enough to produce MeV
mass ALP. Especially, the spectrum peak height does not
decrease much when increasing the ALP mass which indicates
that the projected sensitivity would not have strong
dependence on the ALP mass as long as the kinematic region
is open.

\section{Absorption and Detection at Dark Matter and Neutrino Detectors}
\label{sec:detection}

These MeV solar ALPs produced from the Compton-like process
inside the Sun would come out of the Sun isotropically.
Then the flux arriving at Earth gets diluted by the
sphere surface area which has already been summarized
by the factor $1 / 4 \pi d^2_\odot$ in \geqn{eq:fl}. Being
relativistic, these ALPs can easily overcome the detection
threshold by scattering with the target electrons.

\subsection{Bosonic Absorption Processes}

Being similar as the fermion absorption process
\cite{Dror:2019onn,Dror:2019dib,Ge:2022ius},
an ALP can be absorpted by an atomic electric and convert
its mass to deposit energies in detector. Even better
than its fermionic counterpart, the ALP mass can be 
fully converted through absorption process ($a+e\to e+\gamma$)
that is governed by the same interaction in \geqn{eq:lag}.
Note that the bosonic absorption process is a Compton-like
process.

The detection channel is the inverse of \geqn{eq:prod}
and the Feynman diagram can be realized just by switching 
the incoming $\gamma$ and the outgoing $a$ in \gfig{fig:fd}.
The spin-averaged total amplitude squared takes the form as,
\begin{align}
  \overline{|\mathcal{M}|^2}_{\rm det}
& =
  \frac {e^2 g_{ae}^2}
        {E_{\gamma}^2 m_e \left(m_a^2 + 2 E_a m_e\right)^2}
\bigg\{
  E_{\gamma} m_a^6
+ m_a^4 \left(2 E_a E_{\gamma} + m_a^2\right) m_e
\\
& + 4 (E_a - E_{\gamma}) m_a^2
  \left[ (E_a - E_{\gamma}) E_{\gamma} + m_a^2 \right] m_e^2
+ 4 (E_a - E_{\gamma})^2 \left(2 E_a E_{\gamma} + m_a^2\right) m_e^3
\bigg\}.
\nonumber
\end{align}
Note that the incoming ALP energy $E_a$ and
the outgoing photon energy $E_\gamma$ should not
be confused with their counterparts in the production
process inside the Sun.  

Employing the standard $2\to2$ scattering formalism
in the electron rest frame, the differential
scattering cross section of an incoming ALP with
the target electron reads as,
\begin{eqnarray}
  \frac{d\sigma(E_a)}{dE_\gamma}
=
  \frac 1 {4 m_e |\bm{p}_a|^2}
  \frac 1 {8\pi}
  \overline{|\mathcal{M}|^2}_{\rm det},
  \label{eq:dsigmaC}
\end{eqnarray}
where $\bm{p}_a$ signifies the 3-momentum of the incoming ALP.
The expected differential event rate is a
convolution of the differential cross section
with the incoming ALP flux,
\begin{equation}
  \frac{d R_C}{d E_\gamma}
=
  t_{\rm exp} n_{\rm T} \mathcal{E}(E_\gamma)
  \int^{E_{a}^{\rm max}}_{E_{a}^{\rm min}(E_\gamma)}
  \frac{d\Phi_a}{dE_a}
  \frac{d \sigma_{C}}{dE_\gamma}dE_{a},
\label{eq:recoil}
\end{equation}
where $t_{\rm exp}$ and $n_T$ signify the time of exposure
and the number of target particles, respectively.
For the detector size, PandaX-4T \cite{PandaX:2024cic} and
XENONnT \cite{XENON:2026qow} currently use 3.7 and
5.9 tonnes of liquid xenon as target, respectively. 
In this analysis, we obtain the realistic projected 
limit on $\sim$MeV solar ALP using similar target volumes
with a 1 year run time.
We also discuss the projected limit with an optimistic 200 tonne$\times$year exposure in the future run of  PandaX-xT \cite{PANDA-X:2024dlo}.

In addition, $\mathcal{E}(E_\gamma)$ is the efficiency
factor \cite{PandaX:2024cic,LZ:2025zpw,XENON:2022ltv}
of the experiment and $R_C$ stands for the ALP-$e$
Compton scattering rate.
For experiments with liquid Xenon ($Z=54$) as target
such as PandaX-4T \cite{PandaX:2024cic} and XENONnT
\cite{XENON:2022ltv}, the signal efficiency is highly
suppressed for recoil energy below keV and saturates
to a constant value $\sim 0.95$ and $\sim 1$,
respectively, at energies $\gtrsim \mathcal{O}(10)$ keV. 
Since the energy range ( $E \gtrsim \mathcal{O}(0.1)\mathrm{MeV}$) relevant for our analysis lies well above this threshold, the efficiency is expected to have already saturated and remain approximately constant. Therefore, for the purpose of obtaining a conservative yet simplified estimation of the signal rate, we adopt an optimistic constant efficiency, $\mathcal{E}=1$, throughout our analysis.

Given the signal photon energy $E_\gamma$,
the lower limit of the integration in \geqn{eq:recoil} is determined by 
the minimum energy ($E_a^{\rm min}$),
\begin{eqnarray}
  E_a^{\rm min}
= 
  \frac 1 {2 m_e (2 E_\gamma - m_e)}
\left[
  (E_{\gamma} - m_e)
  \left(2 E_{\gamma} m_e - m_a^2\right)
+ E_{\gamma} \sqrt{\left(m_a^2 + 2 E_{\gamma} m_e\right)^2 - 4 m_a^2 m_e^2}
\right],
\end{eqnarray}
of an incoming ALP required to generate a photon with
energy $E_\gamma$. On the other hand, the integration
upper limit $E^{\rm max}_a$ has been summarized in
\geqn{eq:emin} with $E_\gamma = 5.5$\,MeV.

\begin{figure}[t]
\centering
\includegraphics[width=0.68\textwidth]{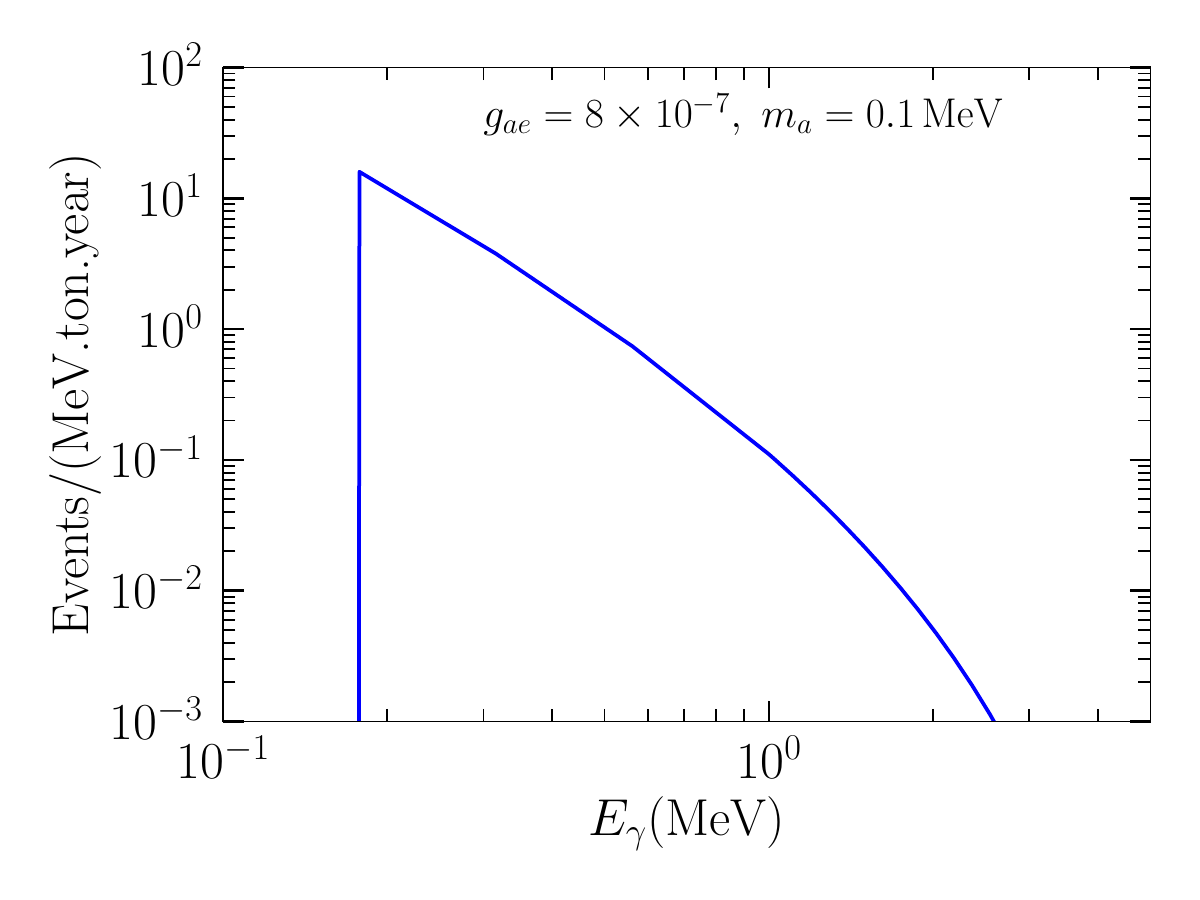}
\caption{The expected event spectra of the recoil
photon energy due to the solar ALP-electron Compton-like
scattering for ALP mass $m_a=0.1$\,MeV and
$g_{ae}=8\times 10^{-7}$.}
\label{fig:spectra}
\end{figure}
\gfig{fig:spectra} displays the event spectrum of
the photon energy $E_\gamma$ following \geqn{eq:recoil}
with a total exposure of 1 tonne$\times$year.
For illustration, these results are obtained with
a benchmark ALP mass $m_a=0.1$\,MeV and coupling
$g_{ae}=8\times10^{-7}$.
The spectrum has a cruel cut around $E_\gamma \sim 0.2$\,MeV
due to kinematics. To be more specific,
the outgoing $\gamma$ energy is restricted as,
\begin{eqnarray}
  E_\gamma^{\rm min}
=
  \frac {m_a^2 + 2 E_a m_e}
        {2 \left( E_a + m_e + \sqrt{E_a^2 - m_a^2} \right)},
\quad
  E_\gamma^{\rm max}
=
  \frac {m_a^2 + 2 E_a m_e}
        {2 \left( E_a + m_e- \sqrt{E_a^2 - m_a^2} \right)}.
\label{eq:egam}
\end{eqnarray}
In the relativistic limit ($m_a\ll E_a$), the minimum
(maximum) energy, $E_\gamma^{\rm min} \to E_a/2$
$(E_\gamma^{\rm max} \to E_a)$, of the outgoing
photon scales with the axion energy $E_a$. 
This has an interesting consequence in the detection
aspect as ALPs with mass beyond the keV scale can
now produce a signal photon with energy $\sim 1$\,MeV,
offering an ideal opportunity of overcoming the detection
threshold. 
Actually, both the photon and electron energies
can deposit in the detector. One may simply take the
incoming axion energy $E_a$ spectrum in \gfig{fig:flux}
as a rough estimation of the measured energy
spectrum. Nevertheless, we still give the
photon energy spectrum in the detector for
completeness.

On the other hand, at higher recoil energies,
the event spectrum exhibits a monotonic decrease.
This suppression is directly correlated with
the rapid decline in the incoming solar ALP
flux at higher energies as shown in \gfig{fig:flux}.
For other values of $m_a$ below 1\,MeV,
the event spectrum does not change significantly
since neither the peak value of the incoming
flux nor the detection cross section have
significant dependence on the ALP mass. 
Consequently, we anticipate that the projected
experimental sensitivity on the coupling will
remain largely the same with respect to the ALP
mass for $m_a\lesssim 1$\,MeV.

Apart from the Compton-like scattering,
there also exists another possible interaction
channel where the ALP knocks out an atomic electron
which is often referred to as the {\it axio-electric
effect} \cite{Avignone:1986vm,Pospelov:2008jk}.
Note that this process is similar to the photo-electric
effect with the incoming photon replaced by an ALP $a$.
The cross section of such a process is given by
\cite{Pospelov:2008jk,Gao:2020wer},
\begin{equation}
  \sigma_{ae}(E_a)
=
  \sigma_{pe}(E_a)
  \frac{g_{ae}^2}{\beta_a}
  \frac{3E_a^2}{16 \pi \alpha m_e^2}
  \left(1-\frac{\beta^{2/3}}{3}\right),
  \label{eq:axioelec}
\end{equation}
where $\sigma_{pe}$ represents the photoelectric
cross section for Xe atom \cite{Veigele:1973aw}
while $\beta=|{\bm p_a}|/E_a$ is the ALP velocity and
$\alpha$ denotes the fine structure constant.
Given this cross section, the event spectrum can be obtained by folding it with the incident ALP flux. Explicitly, the total event rate from the axio-electric process is given by,
\begin{eqnarray}
  R_{ae}
\equiv
  t_{\rm exp} n_{A}
  \int_{E_a^{\rm min}}^{E_a^{\rm max}} \sigma_{ae}
  \frac{d\Phi_a}{dE_a}  dE_a,  
\end{eqnarray}
where $t_{\rm exp}$ is the previously mentioned time of exposure and $n_A$ represents the number of target atoms.

For $E_a\gtrsim 1$\,MeV, the
{\it axion-induced external pair production} process comes into play.
In such a process, the incoming ALP produces an
electron-positron pair in the electric field of Xe nuclei e.g.
$a+\,_{54}^{131}{\rm Xe}\to \,_{54}^{131}{\rm Xe} +e^+ +e^-$
\cite{Kim:1982xb,Kim:1984ss,Blumlein:1991xh}.
The cross section $\sigma_{aee}$ has already been
calculated in Ref.\cite{Arias-Aragon:2024gdz}.
In the massless limit ($m_a\ll E_a$), the cross-section
takes a comparatively simpler form \cite{Arias-Aragon:2024gdz},
\begin{equation}
  \sigma_{aee}
=
  2.6 \frac{g_{ae}^2}{4 \pi \alpha} \sigma_{\gamma e e}(E_a),
\end{equation}
where $\sigma_{\gamma e e}$ represents the photon induced external pair production
cross section for Xe nucleus \cite{xcom_nist}.
The total event rate from such a process is given by,
\begin{eqnarray}
  R_{aee}
\equiv
  t_{\rm exp} n_{A}
  \int_{E_a^{\rm min}}^{E_a^{\rm max}} \sigma_{aee}
  \frac{d\Phi_a}{dE_a}  dE_a. 
\end{eqnarray}

\begin{figure}[t]
\centering
\includegraphics[width=0.68\textwidth]{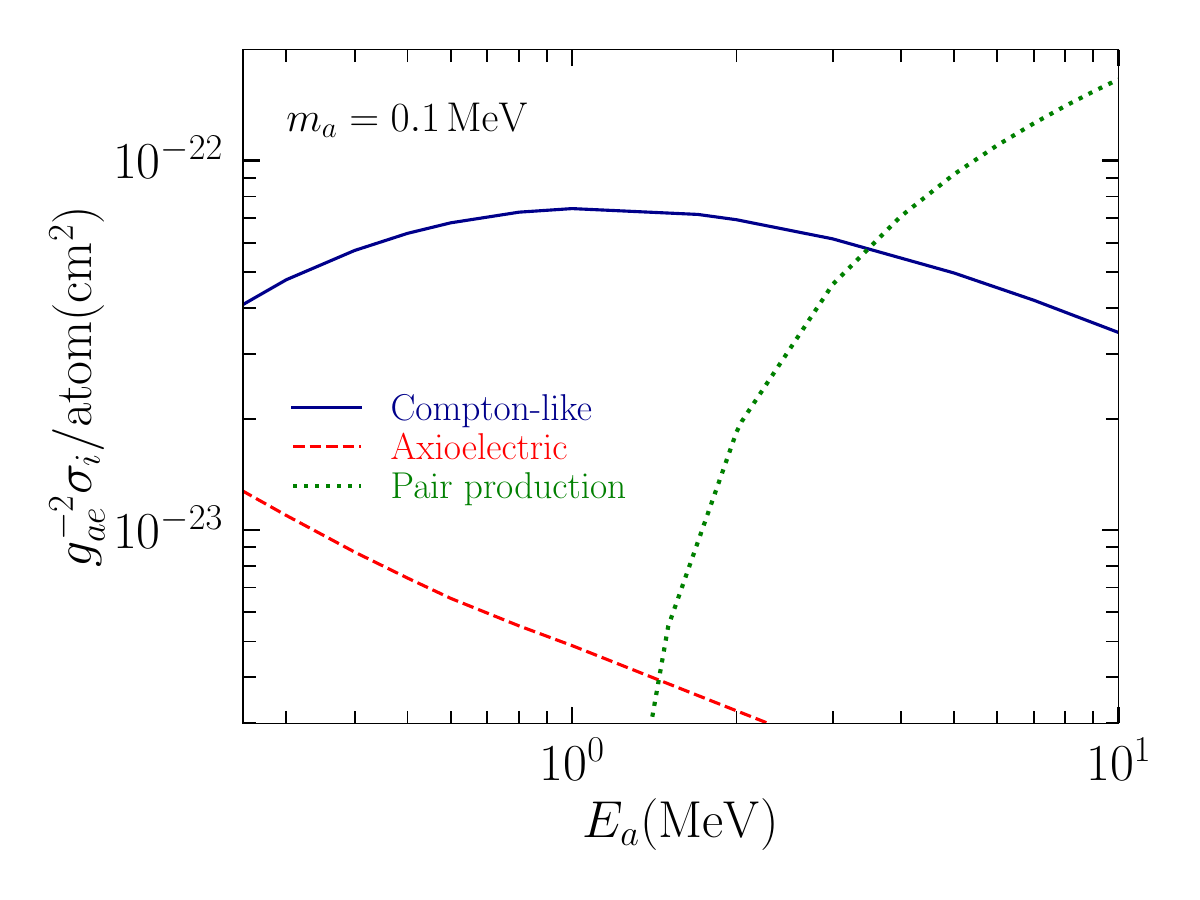}
\caption{Comparison of the ALP-$e$ cross sections per Xenon
atom for the Compton-like scattering
(dark blue), the axio-electric effect (red)
and the axion induced pair production (green)
with benchmark ALP mass $m_a=0.1$\,MeV.
}
\label{fig:comp}
\end{figure}

For comparison, we show in \gfig{fig:comp} the ALP
interaction cross sections per atom for the
Compton-like scattering (dark blue), the axio-electric
effect (red) and axion induced pair-production
(green) with a benchmark ALP mass $m_a=0.1$\,MeV.  
Since the scattering cross sections for all three channels
scale as $\propto g_{ae}^2$, we plot the physical cross
section normalized by $g_{ae}^{-2}$ for a general
discussion.

As evident from the figure, in the energy range 
$0.25\,{\rm MeV}\lesssim E_a\lesssim 5.5$\,MeV of
our interest, the Compton-like scattering dominates
over the axio-electric effect.
This behavior can be understood from the underlying energy dependence of the two processes.
The axio-electric cross section inherits the strong suppression of the photoelectric cross section
$\sim E^{-3}$ \cite{Veigele:1973aw} which leads to
a rapid decrease of $\sigma_{ae}$
with increasing energy despite the explicit $E_a^2$
enhancement in \geqn{eq:axioelec}.
In contrast, the Compton-like scattering process
exhibits a much milder energy dependence in the MeV regime.
As a result, while the axio-electric effect dominates
at keV energies as relevant for the traditional direct
detection experiments \cite{Gao:2020wer,PandaX:2017ock},
its contribution becomes subdominant at higher energies.
Quantitatively, we find that in the MeV energy range, the axio-electric cross section is typically smaller by a factor of a few to an order of magnitude compared with
the Compton-like contribution, depending on the ALP energy.
On the other hand, the axion-induced pair-production process starts to dominate over the Compton-like scattering at $E_a\gtrsim 3$\,MeV.

\subsection{Sensitivities at Dark Matter and Neutrino Experiments}

To place constraints, we evaluate the total event rate
$R_{\rm tot} \equiv R_C+R_{ae}+R_{aee}$ by summing
the contributions from these processes.
However, the suppressed incoming flux at
$E_a\gtrsim 3$\,MeV as shown in \gfig{fig:flux}
leads to a suppressed event rate from the pair
production process for small $m_a(\ll E_a)$.
Consequently, the total event rate $R^{\rm tot}$
is largely driven by the Compton-like scattering
channel in this regime, which eventually determines
the projected sensitivity on the coupling in the
low mass range ($m_a\lesssim 0.5$\,MeV).
However, with the higher values of $m_a\gtrsim0.5$\,MeV,
the axion-induced pair production process is
significantly enhanced and may contribute to the total rate.
The published data from the LZ~\cite{LZ:2019sgr},
PandaX-4T \cite{PandaX:2018wtu}, and Borexino
\cite{BOREXINO:2021efb} experiments are used to constrain
the parameter space of ALPs.

Both LZ~\cite{LZ:2019sgr} and PandaX-4T~\cite{PandaX:2018wtu}
are DM direct detection experiments and are typically sensitive
to $\mathcal O(10)$\,GeV mass DM particles. Although the MeV
energy of the solar ALPs seems much smaller, the electron
recoil energy is much larger and the detection threshold
can be easily overcame. Fortunately, both
LZ \cite{LZ:2022ysc} and PandaX-4T \cite{PandaX:2024sds}
have published their data in the energy range up to 700\,keV
or so.

\begin{figure}[t]
\centering
\includegraphics[width=1.\textwidth]{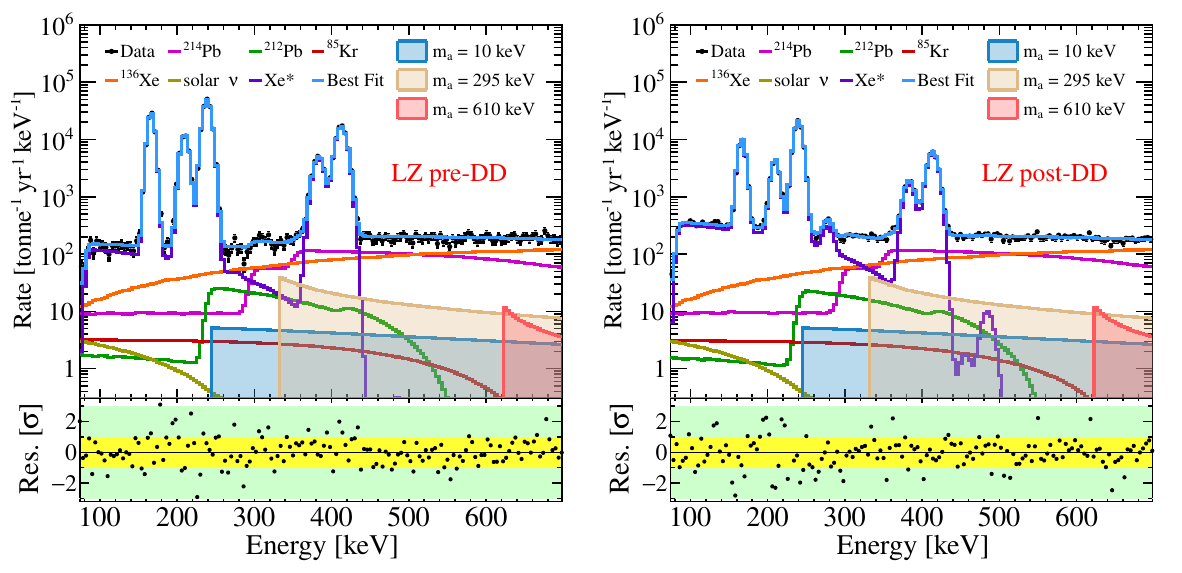}
\caption{The spectrum fitting results of the LZ experiment
with the upper panels showing the event rates \cite{LZ:2022ysc} and the lower
ones for the residual deviations.
While the backgrounds and the total best-fit results are shown
with solid curves, the ALP signals are demonstrated with
filled regions for $m_a = 10$\,keV (blue), 295\,keV (brown),
and 610\,keV (red). The left and right panels are for the
pre-DD and post-DD results, respectively.}
\label{fig:lz_fit}
\end{figure}

\begin{figure}[t]
\centering
\includegraphics[width=0.8\textwidth]{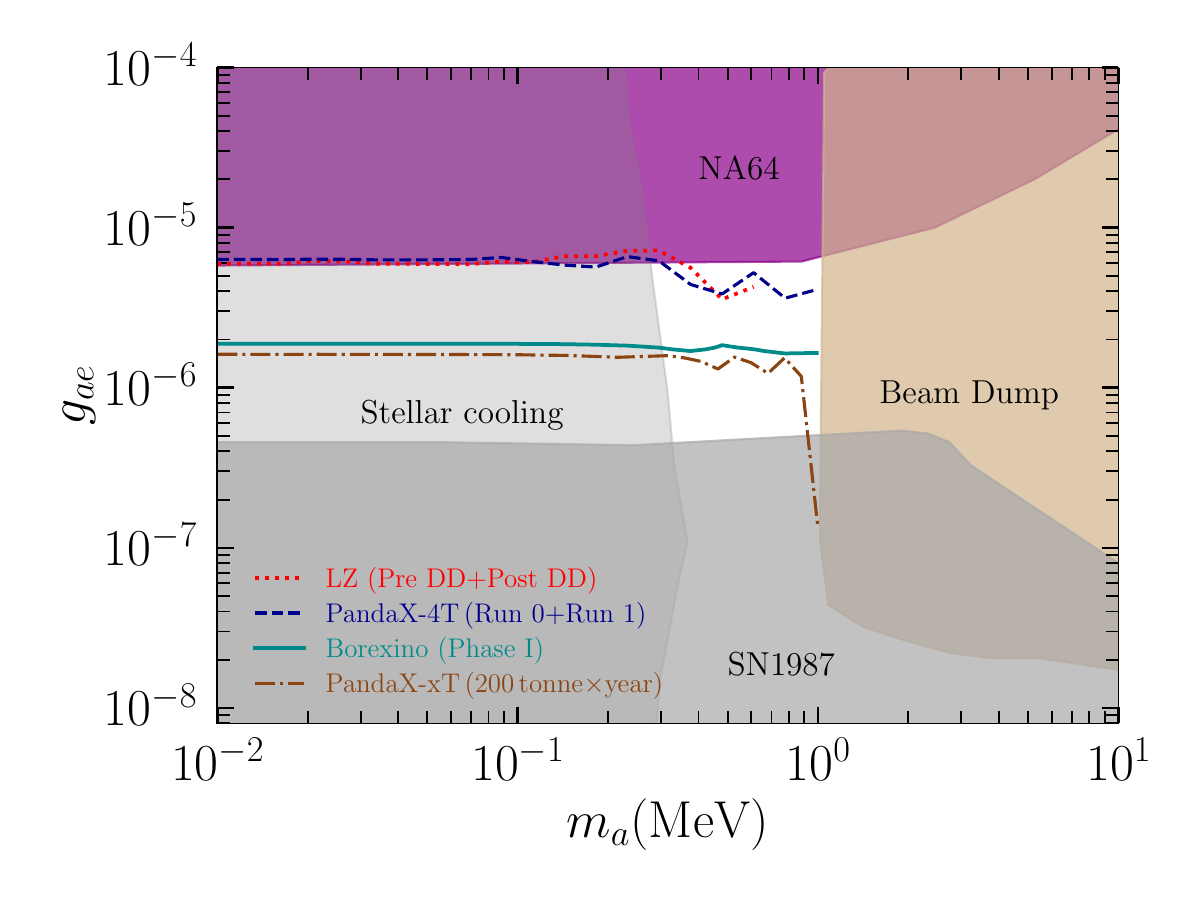}
\caption{The projected $90\%$\,C.L. limit in the $g_{ae}$ vs. $m_a$
plane at the DM and neutrino experiments such as LZ (red dotted),
PandaX-4T (blue dashed), Borexino (dark cyan solid), and PandaX-xT
(brown dot-dashed). The existing constraints are also
portrayed with filled regions for comparison.
}\label{fig:limit}
\end{figure}

The LZ data \cite{LZ:2022ysc} is divided into the pre-DD
and post-DD data sets with exposures of 41 and 124\,kg$\cdot$yr,
respectively. Although the signal energy can extend beyond
1\,MeV, we adopt the same region of interest (ROI) up to
700\,keV \cite{LZ:2022ysc}. The background spectra
before applying the detector response are generated using
a simplified Geant4-based \cite{GEANT4:2002zbu, Allison:2016lfl}
simulation. In addition, the prior constraints on the
various background components are adopted from
Ref.~\cite{LZ:2022ysc}.
A background-only fit is first performed with
$\chi^2_{\rm min}$/NDF of 1.04 and hence the data can
be well described by the background model.
We then perform the profile likelihood scan with the
hypothetical ALP signals in the mass range from 10\,keV
to 1\,MeV. The resulting 90\% confidence level upper
limits are converted into constraints on the ALP
coupling strength as shown in \gfig{fig:limit} with the red
dotted curve.

The PandaX-4T experiment has also published their electron
recoil data \cite{PandaX:2024sds}. With a fiducial mass
of approximately 620\,kg of natural xenon for both Run 0
and Run 1, the total exposure reaches 440\,kg$\cdot$yr.
The ROI is defined as [70, 1000]\,keV, where the lower
boundary excludes the contributions from ${}^{124}$Xe
and ${}^{125}$I,
and the upper boundary as in Ref.~\cite{PandaX:2024sds}. 
We follow the background model as
provided in the Table II of \cite{PandaX:2024sds} which
is mainly contributed by Xe and Pb. The sensitivity
from the PandaX-4T run 0 and run 1 data has
been shown as blue dashed curve in \gfig{fig:limit}.

\begin{figure}[t]
\centering
\includegraphics[width=1.\textwidth]{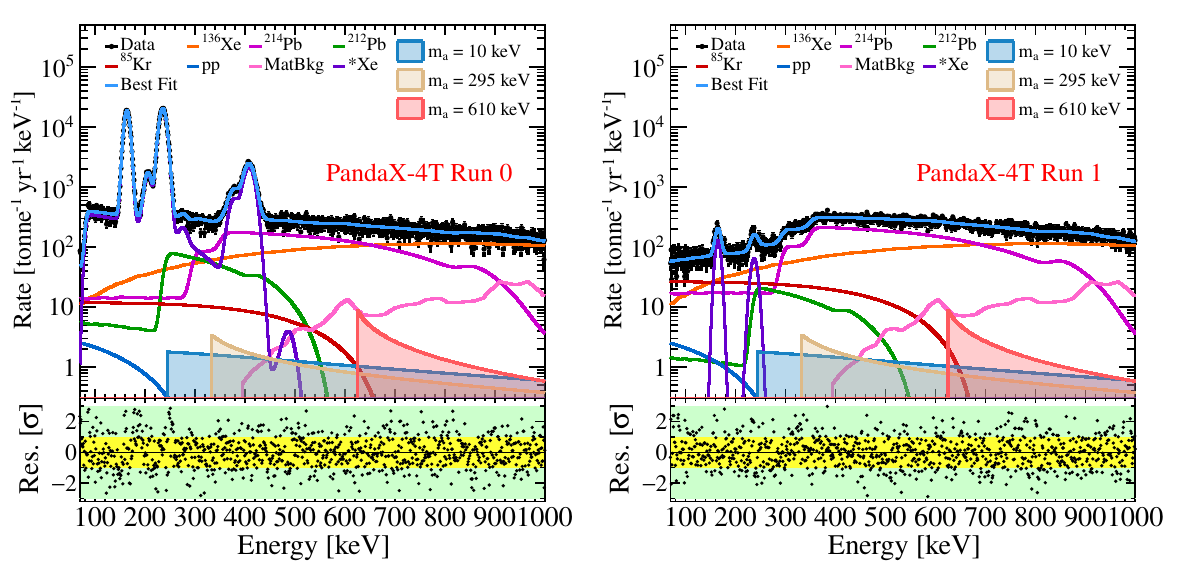}
\caption{The spectrum fitting results of the PandaX experiment
with the upper panels showing the event rates \cite{PandaX:2024sds} and the lower
ones for the residual deviations.
While the backgrounds and the total best-fit results are shown
with solid curves, the ALP signals are demonstrated with
filled regions for $m_a = 10$\,keV (blue), 295\,keV (brown),
and 610\,keV (red). The left and right panels are for the
PandaX-4T Run 0 and Run 1 results, respectively.}
\label{fig:p4_fit}
\end{figure}

For the future PandaX-xT \cite{PANDA-X:2024dlo},
we project its sensitivity with measurements in the
energy range of $0.1 \sim 5.5$\,MeV by using only
the total signal event rate $R^{\rm tot}$ as a
conservative estimation. We evaluate the
projected sensitivity using the Asimov estimate of the significance, $\mathcal{Z} \equiv S/\sqrt{S+B}$ 
\cite{Cowan:2010js}, with $S$ and $B$ being the signal
and background event rates, respectively. 
To place the projected limit, we assume a negligible background ($B=0$) and optimistic $100\%$ detection efficiency across the energy range for the future run of DM detectors.
Under these assumptions, we derive the projected
$90\%$ confidence limit (C.L.) bound on $g_{ae}$
by requiring $\mathcal{Z}=1.64$
\cite{Cowan:2010js,ParticleDataGroup:2024cfk} for a given value of $m_a$.
Note that the total rate $R^{\rm tot}$ scales as $g_{ae}^4$,
allowing a straightforward extraction of the corresponding
sensitivity on the coupling strength $g_{ae}$.
The result has been illustrated with brown dash-dotted
curve in \gfig{fig:limit}.

While the DM direct detection experiments have the benefit
of low energy threshold, the neutrino experiments can compensate
the disadvantage by large volume. We will take the Borexino
experiment for illustration. The Phase-I data \cite{BOREXINO:2021efb}
in the recoil energy range $300 \sim 1600$\,keV has an exposure
of $153.6$\,tonne$\times$year \cite{Borexino:2013zhu}.
Adopting the continuous background spectrum \cite{BOREXINO:2021efb}
and splitting into $10$\,keV bins, we consider the $\chi^2$ fit with,
\begin{equation}
  \chi^2
\equiv
  2 \sum_i
\left[
  R_{\rm pred}^i(m_a,g_{ae})
- R_{\rm exp}^i
+ R_{\rm exp}^i
  \left( \frac{R_{\rm exp}^i}{R_{\rm pred}^i(m_a,g_{ae})} \right)
\right].
\end{equation}
The predicted total event rate $R_{\rm pred}^i(m_a,g_{ae}) \equiv R_{\rm BSM}^i(m_a,g_{ae})+R_{\rm bkg}^i$
in the $i$-th bin is a sum of the ALP
contribution $R_{\rm BSM}^i(m_a,g_{ae})$  for a given
($m_a,g_{ae}$) and the SM background $R_{\rm bkg}^i$.
We obtain the 90\%\,C.L. upper limit on $g_{ae}$ according to
$\Delta \chi^2=\chi^2-\chi^2_{\rm min} = 2.71$ where
$\chi^2_{\rm min}$ is the minimal value of the $\chi^2$ function.
For comparison, the Borexino sensitivity curve is depicted
by a blue solid curve in \gfig{fig:limit}.

We display the obtained limits in the $g_{ae}$ vs. $m_a$
plane in \gfig{fig:limit} for direct and detailed comparisons.
An overall feature is that
the sensitivity becomes largely independent of 
$m_a$ in the low-mass regime as the event rate becomes
insensitive to the ALP mass for $m_a\lesssim 0.1$\,MeV,
leading to a nearly flat exclusion contour. This is especially
true when the mass becomes light enough to have negligible effects.
However, when $m_a$ approaches 1\,MeV, the sensitivity would
significantly improve which can be clearly seen in the
PandaX-xT curve. This is because the electron-positron
pair production rate significantly increases when approaching
the double electron mass $2 m_e \sim 1$\,MeV. However,
the existing data from the DM direct detection experiment
LZ (PandaX-4T)  considered here only covers the energy range up to 700\,keV (1000\,keV). 
Although the PandaX-4T and Borexino data can extend beyond 1\,MeV,
the uncertainties there is also much larger to weaken the expected sensitivity enhancement.

Among these experiments, LZ and PandaX-4T can already probe
couplings as small as $g_{ae}\gtrsim 3.7 \times 10^{-6}$ 
for $m_a\lesssim 0.1$\,MeV. With larger exposure, Borexino
pushes the constraint to $g_{ae}\gtrsim 1.7 \times 10^{-6}$
for $m_a\lesssim 1$\,MeV. On the other hand, an optimistic
200 tonne$\times$year exposure of PandaX-xT can further
touch down to $g_{ae}\gtrsim 1.5 \times 10^{-7}$ for $m_a\lesssim1$\,MeV.
The slight improvement in the PandaX-xT projection for $m_a\gtrsim 0.7$\,MeV is due to the enhancement in the rate caused by ALP-induced pair production.
However, such an effect is not prominent in other limits due to the high backgrounds and smaller recoil energy range considered. 
The improvement in sensitivity with increasing exposure follows from the scaling of the significance with the square root of the number of signal events ($\propto t_{\rm exp}n_T g_{ae}^4$), resulting in a stronger reach for larger detector exposures.

In the same plane of \gfig{fig:limit},
we also show the existing constraints
on the electrophilic ALP. In the range $m_a\lesssim 2m_e$,
one of the most stringent experimental constraints comes
from NA64  which is a fixed-target experiment that uses
electron beam with energy $E_e=$100\,GeV imparted on
a fixed proton target to produce the scalar ALP.
This gives
$g_{ae}\lesssim 10^{-5}$ \cite{NA64:2020qwq,NA64:2021aiq}
that is depicted by the magenta shaded region.
For heavier ALPs with mass $m_a\gtrsim 2m_e$, those beam
dump experiments like E137, E141, and Orsay
\cite{Bjorken:1988as,Bechis:1979kp} provide the leading
constraints. These experiments look for the ALP visible
decay through the displaced vertex search 
and provide a bound $g_{ae}\lesssim 5\times 10^{-7}$ for  $1$\,MeV $\lesssim m_a\lesssim100$\,MeV \cite{Blumlein:1991xh,Andreas:2010ms,Liu:2017htz,Waites:2022tov} as shown by the light brown filled region.

In the low-mass regime, the astrophysical observations provide some of the most stringent constraints on the electrophilic ALPs. In particular, the stellar cooling
arguments place strong bounds on light ALPs with feeble couplings. Such particles can be efficiently produced inside stellars such as red giants and
white dwarfs etc. before escaping the stellar core
without further interactions to provide an additional
energy loss channel \cite{Giannotti:2015kwo}. Such
stellar cooling constraints are shown by the light grey
shaded region.
The supernova cooling considerations provide complementary
constraints in a similar mass range. In the core of a
supernova such as SN1987A, ALPs can be produced in the
hot and dense environment and escape, thereby carrying
away energy.
Requiring consistency with the SN1987A neutrino burst leads to
bounds on the coupling, typically relevant for $g_{ae}\lesssim 5\times 10^{-7}$
and $m_a\lesssim \mathcal{O}(1)$\,MeV depicted by the dark
grey shaded region \cite{Carenza:2021pcm,Ferreira:2022xlw}.
We would like to mention that these astrophysical constraints
dominate over those laboratory limits in the low-mass region,
although they are subject to larger uncertainties associated
with the stellar modeling \cite{Bar:2019ifz}.

The ALPs produced in the Sun may, in principle, undergo scattering with the solar electrons to cause attenuation of the outgoing flux.
The mean free path $l \equiv (n_e\sigma_C)^{-1}$ of the
produced ALPs in the sun scales inversely
with the average solar electron density $n_e$ and the
Compton-like scattering cross section $\sigma_C$
between ALP and electron as summarized in \geqn{eq:dsigmaC}.
For, $m_a\ll E_a\sim$\,MeV , the cross-section reads as $\sigma_C\sim 10^{-25}g_{ae}^2{\rm cm}^2$.
Hence, for  $g_{ae}< 10^{-5}$, the mean free path ($l\gtrsim 10^6$\,km)
of the produced ALP exceeds the
solar radius, ensuring that they escape the Sun without significant attenuation. 
Since the region with $g_{ae}\gtrsim 10^{-5}$ is already excluded from existing laboratory constraints like NA64, we neglect the attenuation effect for the analysis in our current paper.

Despite all the stringent constraints, our projected sensitivity demonstrates that the future run of DM detectors can probe previously unexplored regions of the parameter space.
In particular, we identify a window in the mass range
$0.3\,{\rm MeV}\lesssim m_a\lesssim 1$\,MeV where a null observation can lead to the strongest constraint in that range.
This highlights the potential of next-generation detectors to
explore ALP parameter space that is otherwise accessible only indirectly.
We therefore propose a novel strategy to directly probe the
energetic electrophilic ALPs of solar origin and provide a new
avenue to test this class of models in the upcoming run of DM
detection experiments.

\section{Conclusion}
\label{sec:concl}
In this work, we have explored the solar production of
MeV scale electrophilic ALP using the 5.5\,MeV photon
from the $pp$ chain that scatters with the ambient electrons.
Since the typical momentum transfer is at just the MeV
scale, we adopt an EFT framework where the ALP ($a$) couples
with electron through a Yukawa-like interaction. While
the previous studies consider only the nucleophilic ALP,
our paper points out the possibility of using the electrophilic
interactions to produce an MeV ALP.

Despite the monochromatic nature of the incident photons ($E_\gamma\sim5.5$\,MeV), the emitted ALPs exhibit a continuous energy spectrum governed by the $2\to2$
scattering kinematics.
In particular, the energy-momentum conservation poses a sharp cut off at the minimum energy of the produced ALP, $E_a^{\rm min} \approx 0.25$\,MeV for a vanishingly small ALP mass, $m_a \ll m_e \ll E_\gamma$.
Furthermore, the height of the spectrum peak is largely insensitive to the ALP mass in the range, $m_a\lesssim1$\,MeV.
Consequently, the signal event rate and sensitivity are also expected to exhibit a negligible dependence on the ALP mass in that regime.

Upon reaching the Earth, such MeV scale ALPs can scatter off the target electrons in the DM direct detection experiments such as PandaX-4T and LZ via the process $a+e\to e+\gamma$, leading to a distinct recoil spectrum of the outgoing $\gamma$.
Again, the recoil photon energy spectrum has a cruel cut
around the minimum energy $E_\gamma\sim 0.2$\,MeV for an ALP with a vanishingly small mass $m_a\lesssim E_a$
as dictated by the kinematics.
On the other hand, the maximum energy of the recoil photon can reach up to a few MeV.
The event spectrum is also found to be largely invariant with ALP mass in the regime $m_a \lesssim 1$\,MeV.

In addition to the Compton-like scattering, the incoming
ALP may also undergo the {\it axio-electric} process with
the target electrons in the detector. 
However, we observe that the corresponding axio-electric cross section exhibits a rapid decrease with increasing $E_a$, rendering this channel subdominant comparing with
the Compton-like process in the energy range 
$0.25\,{\rm MeV}\lesssim E_a\lesssim 5.5$\,MeV that is
of our interest.
As a result, the signal event rate is largely driven by the Compton-like scattering up to $m_a\lesssim 0.5$\,MeV.
However, for $m_a\gtrsim 0.5$\,MeV the ALP-induced pair
production process also starts to contribute to the total
rate, leading to a slight improvement in the sensitivity
in that mass regime.

We obtain the $90\%$\,C.L. limits on the $m_a$ vs. $g_{ae}$
parameter space considering LZ, PandaX-4T
and Borexino experiments. Our results indicate that
the current sensitivities from LZ, PandaX-4T and Borexino
can reach $g_{ae}\gtrsim 3.7 \times 10^{-6}$,
$g_{ae}\gtrsim 3.7 \times 10^{-6}$ and
$g_{ae}\gtrsim 1.7 \times 10^{-6}$,
respectively, for $m_a\lesssim1$\,MeV.
An upgraded configuration, such as PandaX-xT with an optimistic exposure of 200 tonne$\times$year has the potential to reach  $g_{ae}\lesssim 1.6 \times 10^{-6}$ for
most of the mass window $m_a < 1$\,MeV and even
$g_{ae}\lesssim 1.5\times 10^{-7}$ for $m_a$ approaching 1\,MeV.
As expected from the behavior of the ALP flux and the corresponding recoil spectra, the projected experimental sensitivities on the coupling remain largely the same with respect to the ALP mass for $m_a\lesssim 1$\,MeV.
Despite the stringent existing constraints from different laboratory experiments and astrophysical observations on the same parameter space, our projected limits extend into previously unexplored regions of parameter space, particularly in the mass window $m_a \in (0.3 \sim 1)$\,MeV.
For smaller ALP mass ranges ($m_a\lesssim 0.3$\,MeV), our projections are complementary to the existing constraints from stellar cooling  and can provide the first direct probe in that particular parameter space.

\section*{Acknowledgements}
The authors would like to thank Xue-Feng Ding, Ruofei Feng, Ke Han, Jianglai Liu,
Jo\~ao Paulo Pinheiro, Oleg Titov, Yongchao Zhang for useful discussions and various suggestions.
This work is supported by the National Natural Science
Foundation of China (12375101 and 12425506), by the Office of Science and Technology, Shanghai Municipal Government (25ZR1402223), and by the Natural Science Foundation of Sichuan, China (2026NSFSC0757).
SFG is also an affiliate member of Kavli IPMU, University of Tokyo.
This work is also supported by State Key Laboratory of Dark Matter Physics.

\appendix

\bibliography{ref}
\bibliographystyle{utphysGe}

\end{document}